\documentclass[aps,prd,onecolumn,showpacs,showkeys,amsmath,amssymb]{revtex4}
\usepackage{graphicx}
\usepackage{dcolumn}
\usepackage{bm}
\begin{document}
\title{Pionic Decays of $D_{sj}(2317)$, $D_{sj}(2460)$
and $B_{sj}(5718)$, $B_{sj}(5765)$}
\author{Jie Lu}
\affiliation{Department of Physics, Peking University, Beijing
100871, China}
\author{Wei-Zheng Deng}
\email{dwz@th.phy.pku.edu.cn} \affiliation{Department of Physics,
Peking University, Beijing 100871, China}
\author{Xiao-Lin Chen}
\affiliation{Department of Physics, Peking
University, Beijing 100871, China}
\author{Shi-Lin Zhu}
\email{zhusl@th.phy.pku.edu.cn} \affiliation{Department of
Physics, Peking University, Beijing 100871, China}

\date{\today}

\begin{abstract}

We estimate pionic decay widths of the narrow charm-strange
resonances $D_{sj}(2317)$ and $D_{sj}(2460)$ using the $^3P_0$
model. Their one-pion decays occur through $\eta$-$\pi^0$ mixing
while the two-pion decays of $D_{sj}(2460)$ occur through the
virtual $f_0(980)$ meson. The mixing between $^3P_1$ and $^1P_1$
states enhances the single pion decay width of $D_{sj}(2460)$ and
suppresses its double pion decay width significantly. The two-pion
decay width of $D_{sj}(2460)$ is much smaller than its one-pion
decay width. As a byproduct, we also calculate pionic decay widths
of $B_{sj}(5765)$, $B_{sj}(5765)$ mesons in the $(0^+, 1^+)$ heavy
doublet.

\end{abstract}



\maketitle

\pagenumbering{arabic}

\section{Introduction}\label{sec1}

In April 2003, Babar Collaboration reported a new charm-strange
state $D_{sj}(2317)$ in the $D_s\pi^0$ channel. Its spin-parity is
$J^P=0^+$ and its mass is below the $DK$ threshold 2.36 GeV. Later
CLEO Collaboration reported another new charm-strange resonance
$D_{sj}(2460)$ in the $D^*_s\pi^0$ channel with $J^P=1^+$ below
the $D^*K$ threshold \cite{Besson:2003cp}. It's tempting to
classify these two states as the $(0^+, 1^+)$ P-wave $c\bar{s}$
doublet. But their observed masses are more than one hundred MeV
lower than quark model predictions \cite{quark model}. There have
been heated debates about its underlying structure and origin of
its low mass in the literature
\cite{Bardeen,Nowak,lutz,molecule,tetra1,tetra2,tetra3,tetra4,tetra5,
tetra6,tetra7,tetra8,atom,deandrea,cahn,slz,lucha,hofmann,sadzi,
beci,lee,wei,alz,lattice,colangelo4,liu}. Although several
non-conventional schemes such as $DK$ molecules \cite{molecule},
four quark states
\cite{tetra1,tetra2,tetra3,tetra4,tetra5,tetra6,tetra7,tetra8} and
$D\pi$ atom \cite{atom} were proposed, evidence is gradually
accumulating that these two narrow resonances are ordinary $c\bar
s$ states
\cite{Bardeen,Nowak,deandrea,cahn,slz,lucha,hofmann,sadzi,beci,lee,wei}.

Another interesting property of $D_{sj}(2317)$ and $D_{sj}(2460)$
mesons is their extremely narrow widths. The decay channels
$D_{sj}(2317)\rightarrow D\;K$ and $D_{sj}(2460)\rightarrow
D^*\;K$ are forbidden by kinematics. Therefore, their possible
strong decay modes are one-pion and two-pion decays. The two-pion
decay occurs via a virtual meson such as $f_0(980)$. The one-pion
decay mode breaks the isospin symmetry and happens through
$\eta$-$\pi^0$ mixing \cite{wise}: $D_{sj}(2317)\rightarrow
D_s\eta\rightarrow D_s\pi^0$, $D_{sj}(2460)\rightarrow
D^*_s\eta\rightarrow D^*_s\pi^0$. The $\eta-\pi^0$ mixing is
described by the isospin violating piece in the chiral lagrangian
\begin{equation}
{\cal{L}}_m=\frac{m_{\pi}^2 f^2}{4(m_u+m_d)}\mbox{Tr}(\xi m_q
\xi+\xi^{\dagger} m_q \xi^{\dagger})~,
\end{equation}
where $\xi= \exp (i\tilde{\pi}/f_\pi)$,  $\tilde{\pi}$ the light
meson octet and $m_q$ is the light quark mass matrix. Such a
mixing is suppressed by the factor
$\frac{m_d-m_u}{m_s-\frac{m_u+m_d}{2}}$. Numerically the isospin
violating effect is ${\cal O}(10^{-2})$ in the amplitude. While
the isospin conserving strong decay width is ${\cal O}(10^{2})$
MeV, one would naturally expect the one-pion decay width of
$D_{sj}(2317)$ and $D_{sj}(2460)$ to be around several tens keV.

Although quantumn chromodynamics (QCD) is widely accepted to be
the correct theory of strong interaction, our present
understanding of the strong decay mechanism is rather limited.
First-principle calculation of decay matrix elements on the
lattice still has a long way to go. For some limited special
cases, complicated tools such as light-cone QCD sum rules may be
used to calculate the coupling constant \cite{wei}. In order to
understand the vast amount of strong decay data, we have to turn
to phenomenological strong decay models. Among them, the $^3P_0$
model (or quark pair creation model) is the simplest and most
successful one. In this work, we employ this model to calculate
the decay amplitude of $D_{sj}(2317)$ and $D_{sj}(2460)$ mesons
assuming they are $c\bar{s}$ states. A short review of $^3P_0$
model is given in Section \ref{sec2}. The analysis of the one-pion
decays is presented in Section \ref{sec3}. The two-pion decay mode
of $D_{sj}(2460)$ is given in Section \ref{sec4}. The last section
is a short summary.

\section{The $^3P_0$ model }\label{sec2}

\subsection{The model}

\begin{figure}
\begin{center}
\scalebox{0.6}{\includegraphics{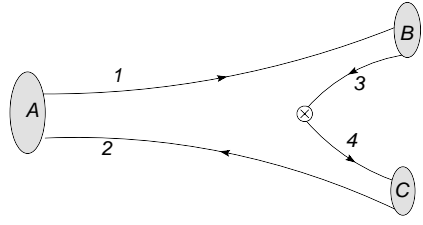}}
\end{center}
\caption{The meson decay $A \rightarrow B C$ in the $^3P_0$ model.
}\label{fig1}
\end{figure}

The $^3P_0$ model was first introduced by Micu in 1969
\cite{Micu69} and further developed by the Orsay group in the
1970s \cite{orsay70}. According to this model, a $q\bar q$ pair
with $J^{PC}=0^{++}$ is created from the vacuum when a hadron
decays. This created $q\bar q$ pair carries the quantum number of
the vacuum. The new $q\bar{q}$ pair, together with the $q\bar{q}$
within the parent meson regroups into the outgoing mesons via
quark rearrangement process, which is shown pictorially in Fig.
\ref{fig1}.
The transition operator in the nonrelativistic limit reads
\cite{orsay70,Godfery96}
\begin{equation}
T = - 3 \gamma \sum_m\: \langle 1\;m;1\;-m|0\;0 \rangle\,
(2\pi)^{3\over2}\int\!{\rm d}^3{\textbf{k}}_3\; {\rm
d}^3{\textbf{k}}_4 \delta^3({\textbf{k}}_3+{\textbf{k}}_4)\: {\cal
Y}^m_1({\textbf{k}}_3-{\textbf{k}_4})\; \chi^{3 4}_{1 -\!m}\;
\phi^{3 4}_0\;\, \omega^{3 4}_0\; b^\dagger_3({\textbf{k}}_3)\;
d^\dagger_4({\textbf{k}}_4) \label{tmatrix}
\end{equation}
where $\gamma$ is the dimensionless pair creation parameter in the
model, and indicates the strength of coupling. ${\cal
Y}^m_l({\bf{k}}) \equiv k^l\;
Y^m_l(\theta_k,\phi_k)=-\epsilon_m\cdot {\bf{k}}$ is a solid
spherical harmonic function that describes the momentum
distribution of the created pair. The meson state is defined as
\cite{mockmeson}:
\begin{eqnarray}\label{mockmeson}
|A(n_A \mbox{}^{2S_A+1}L_A \,\mbox{}_{J_A M_{J_A}})
({\textbf{p}}_A) \rangle &=& \sqrt{2 E_A}\:
\!\!\!\!\!\!\sum_{M_{L_A},M_{S_A}}\!\!\! \langle L_A M_{L_A}
S_A M_{S_A} | J_A M_{J_A} \rangle \nonumber\\
&&\!\!\!\!\!\!\!\!\!\!\!\!\times \;\!\!\int\!{\rm
d}^3{\textbf{k}}_1{\rm
d}^3{\textbf{k}}_2\delta^3({\bf{k_1+k_2\!-\!p_A}})\psi_{n_A L_A
M_{L_A}}\!({\bf{k_1,k_2}})\chi^{1 2}_{S_A M_{S_A}}\phi^{1
2}_A\omega^{1 2}_A |\;q_1({\textbf{k}}_1)
\bar{q}_2({\textbf{k}}_2)\rangle  \nonumber\\
\langle A({\textbf{p}}_A)|A({\textbf{p}}'_A) \rangle &=&
2E_A(2\pi)^3\delta^3({\textbf{p}}_A-{\textbf{p}}'_A)\; .
\end{eqnarray}
The subscripts 1 and 2 refer to the quark and antiquark within the
parent meson A respectively. ${\textbf{k}}_1$ and ${\textbf{k}}_2$
are the momentum of the quark and antiquark, ${\textbf{k}}_A$ is
their relative momentum and ${\textbf{p}}_A$ is the momentum of
meson A. $S_A=s_{q_1}+s_{q_2}$ is the total spin. $J_A=L_A+S_A$ is
the total angular momentum.

The S-matrix reads
\begin{eqnarray}
\langle
f|S|i\rangle&=&I+i(2\pi)^4\delta^4(p_f-p_i)\mathcal{M}^{M_{J_A}
M_{J_B} M_{J_C}}\;.
\end{eqnarray}
The decay helicity amplitude of the process $A\rightarrow B+C$ in
the meson A center of mass frame is
\begin{eqnarray}
&&{\mathcal{M}}^{M_{J_A} M_{J_B} M_{J_C}}(A\rightarrow BC)
\nonumber\\ &=&\sqrt{8 E_A E_B
E_C}\;\;\gamma\!\!\!\!\!\!\!\!\!\!\!
\sum_{\renewcommand{\arraystretch}{.5}\begin{array}[t]{l}
\scriptstyle M_{L_A},M_{S_A},\\
\scriptstyle M_{L_B},M_{S_B},\\
\scriptstyle M_{L_C},M_{S_C},m
\end{array}}\renewcommand{\arraystretch}{1}\!\!\!\!\!\!\!\!
\langle L_A M_{L_A} S_A M_{S_A} | J_A M_{J_A} \rangle  \langle L_B
M_{L_B} S_B M_{S_B} | J_B M_{J_B} \rangle \langle L_C M_{L_C} S_C
M_{S_C} | J_C M_{J_C} \rangle \nonumber\\
&& \times  \langle 1\;m;1\;-m|\;0\;0 \rangle\; \langle \chi^{3
2}_{S_C M_{S_C}}\chi^{1 4}_{S_B M_{S_B}}  | \chi^{1 2}_{S_A
M_{S_A}} \chi^{3 4}_{1 -\!m} \rangle
 \langle\varphi^{3 2}_C \varphi^{1 4}_B | \varphi^{1 2}_A \varphi^{3 4}_0
 \rangle
\;I^{M_{L_A},m}_{M_{L_B},M_{L_C}}({\textbf{p}}) \;,\nonumber\\
\end{eqnarray}
where the spatial integral $I_m(A,BC)$ is defined as
\begin{eqnarray}
I^{M_{L_A},m}_{M_{L_B},M_{L_C}}({\textbf{p}})&=&\;(2\pi)^{3\over2}
\int\!{\rm d}^3{\textbf{k}}_1{\rm d}^3{\textbf{k}}_2{\rm
d}^3{\textbf{k}}_3{\rm d}^3{\textbf{k}}_4\;
\delta^3({\bf{k_3+k_4}})\delta^3({\bf{k_1+k_2-p_A}})\delta^3({\bf{k_1+k_3-p_B}})\delta^3({\bf{k_2+k_4-p_C}})
\nonumber\\
&&\times \psi^*_{n_B L_B M_{L_B}}\!
({\textbf{k}}_1,{\textbf{k}}_3)\; \psi^*_{n_C L_C M_{L_C}}\!
({\textbf{k}}_2,{\textbf{k}}_4)\; \psi_{n_A L_A M_{L_A}}\!
({\textbf{k}}_1,{\textbf{k}}_2)\; {\cal Y}^m_1({\bf{k_3-k_4}})
\label{integral}
\end{eqnarray}
We employ the harmonic-oscillator wave functions for S-wave states
\cite{Isgur85}:
\begin{eqnarray} \psi^{L=0}({\bf{k_1,k_2}}) &=& \Big{(}\frac{R^2}{
\pi }\Big{)}^{3/4} \exp\Big{[}-{1\over
8}({\bf{k_1-k_2})}^2R_B^2\Big{]}
\end{eqnarray}
where the parameter $R$ is the meson radius. For the P-wave state,
we use \cite{Isgur85}

\begin{eqnarray}
\psi^{L=1}({\bf{k_1,k_2}}) &=& i\sqrt{2\over 3}{R^{5\over 2}\over
\pi^{1/4}} {\mathcal{Y}}_{1m}({\bf{k_1-k_2}})\exp\Big{[}-{1\over
8}R^2({\bf{k_1-k_2})}^2\Big{]} \;.
\end{eqnarray}

The spin matrix element can be written in terms of Wigner's 9j
symbol \cite{orsay70}
\begin{eqnarray}
&&{ \langle  \chi^{3 2}_{S_C M_{S_C}} \chi^{1 4}_{S_B M_{S_B}}|
\chi^{1 2}_{S_A M_{S_A}} \chi^{3 4}_{1 -\!m} \rangle} \nonumber\\&
=& \sum_{S_{BC},M_s}\langle  S_C M_{S_C};S_B M_{S_B} |S\;M_s
\rangle \langle S\;M_s| S_A M_{S_A};1 -m \rangle\;
\Big{[}3(2S_B+1)(2S_C+1)(2S_A+1)\Big{]}^{1/2} \left
\{\begin{array}{ccc}
1\over 2 & 1\over 2 & S_C \nonumber\\
1\over 2 & 1\over 2 & S_B\\
S_A & 1 & S
\end{array}
\right \} \;.
\end{eqnarray}
The relevant flavor matrix element for $D_{sj}(2317)\rightarrow
D_s\eta$ is
\begin{eqnarray}
\langle \varphi^{3 2}_C \varphi^{1 4}_B| \varphi^{1 2}_A
\varphi^{3 4}_0  \rangle = \Big{\langle} {1\over
\sqrt{3}}\{u\bar{u}+d\bar{d}+s_3\bar{s_4}\}\{c_1\bar{s_2}\}
\Big{|} \{c_1\bar{s_3}\} {1\over
\sqrt{6}}\{u\bar{u}+d\bar{d}-2s_4\bar{s_2}\} \Big{\rangle} =
{2\over \sqrt{18}} \;.
\end{eqnarray}

With the Jacob-Wick formula the helicity amplitude can be
converted into the partial wave amplitude \cite{convert}:
\begin{eqnarray}
{\mathcal{M}}^{J L}(A\rightarrow BC) = \frac{\sqrt{2 L+1}}{2 J_A
+1} \!\! \sum_{M_{J_B},M_{J_C}} \langle L 0 J M_{J_A}|J_A
M_{J_A}\rangle \langle J_B M_{J_B} J_C M_{J_C} | J M_{J_A} \rangle
\mathcal{M}^{M_{J_A} M_{J_B} M_{J_C}}({\textbf{k}}_B)
\end{eqnarray}
where $M_{J_A}=M_{J_B}+M_{J_C}$,
${\textbf{J}}={\textbf{J}}_B+{\textbf{J}}_C $. The decay width in
terms of partial wave amplitudes using the relativistic phase
space is:
\begin{eqnarray*}
\Gamma = {1\over8\pi} \frac{{\textbf{p}}}{M_A^2}\sum_{JL}
|\mathcal{M}^{J L}|^2.
\end{eqnarray*}

For the three body process $A\to BCD$, the width is
\begin{eqnarray}
\Gamma &=&{1\over(2\pi)^5}\;{1\over16M^2_A}\;\int
|\mathcal{M}|^2\; |\mathbf{p}_B|\;|\mathbf{p}^*_C|\; dm_{CD}\;
d\Omega_B \;d \Omega^*_C \;.
\end{eqnarray}

\subsection{ Input Parameters}

Our convention follows Ref.\cite{Godfery96} where the value of
$\gamma =6.9$ is $\sqrt{96\pi}$ times larger than that used by
others groups \cite{close,kokoski}. The masses of $\eta$,
$f_0(980)$ and charm-strange mesons are taken from PDG \cite{pdg}.
Since the $(0^+, 1^+)$ beauty-strange mesons have not been
discovered yet, we use the theoretical estimate of their masses
from Ref. \cite{Bardeen} in our calculation. For the $(0^+, 1^+)$
$D_{sj}$ and $B_{sj}$ mesons, we use the same parameter $R$ from
Ref. \cite{close}. We collect all these values in Table
\ref{tab1}.

\begin{table}[h]
\caption{\baselineskip 15pt Value of mass and R in our calculation
.}\label{tab1}
\begin{center}
\begin{tabular}{c c c c c c c c c c c c c}
\hline
 & $D_s(0^-)$ & $D^*_s(1^-)$ & $D_{sJ}(2317)$ & $D^{^1P_1}_{sJ}(2460)$ &$D^{^3P_1}_{sJ}(2460)$
 & $B_s(0^-)$ & $B^*_s(1^-)$ & $B_s(0^+)$    & $B^{^1P_1}_s(1^+)$ & $B^{^3P_1}_s(1^+)$  &$\eta$ &$f_0(980)$
  \\
\hline mass\;(GeV) & 1.968 &  2.112   & 2.317 & 2.460 & 2.460 &
5.370 &5.417 & 5.718& 5.765    & 5.765  &  0.548  & 0.980
\\ R\;($\mathrm{GeV^{-1}}$)
& 1.37 & 1.59 & 1.89 &1.85 & 1.89  & 1.37 & 1.59 & 1.89 &1.85 & 1.89  &1.47&2.7\\
\hline
\end{tabular}
\end{center}
\end{table}

\section{Single pion decays of $D_{sj}(2317)$ and $D_{sj}(2460)$}
\label{sec3}

\begin{figure}[hbt]
\begin{center}
\scalebox{0.8}{\includegraphics{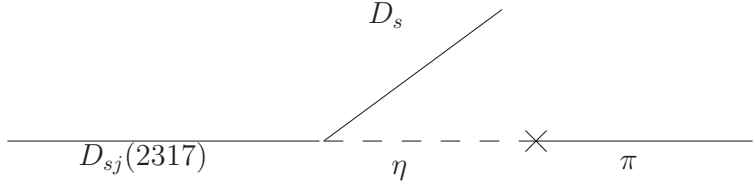}}
\end{center}
\caption{Pionic decay of $D_{sj}(2317)$ via $\eta-\pi^0$ mixing.
}\label{fig2}
\end{figure}

$D_{sj}(2317)$ and $D_{sj}(2460)$ lie below the $DK$ and $D^*K$
thresholds. They can only have pionic decay modes. The
isospin-violating single pion decay occurs through $\eta-\pi^0$
mixing as shown in Fig. \ref{fig2}.

The decay amplitude of $D_{sj}(2317)\rightarrow D_s+\pi^0 $ is
\begin{eqnarray}
{\mathcal{M}}^{J L}\Big{(}D_{sj}(2317)\rightarrow D_s +
\pi^0\Big{)}= {\mathcal{V}}^{J L}\Big{(}D_{sj}(2317)\rightarrow
D_s + \eta\Big{)} \cdot {i\over m_\pi^2-m^2_\eta}\cdot
{\mathcal{V}}(\eta-\pi^0)\;.
\end{eqnarray}

The the ${\mathcal{V}}^{J L}\Big{(}D_{sj}(2317)\rightarrow D_s +
\eta\Big{)}$ term is can be calculated using $^3P_0$ model. The
$\eta-\pi^0$ mixing is determined by the up down quark mass
difference \cite{wise}:
\begin{eqnarray*}
{i\over m_\pi^2-m^2_\eta}\cdot{\mathcal{V}}(\eta-\pi^0)&=&
i{\sqrt{3}\over4} {m_d-m_u\over m_s-{m_u+m_d\over
2}}=i{\sqrt{3}\over4} \delta_{\pi^0\eta}
\end{eqnarray*}
which vanishes in the limit of exact isospin symmetry. The isospin
violating factor $\delta_{\pi^0\eta}={1\over43.7}$ \cite{pdg}.

 In the heavy quark limit, there is one $1^+$ state with
$j_l=L+s_q={1\over 2}^+$ in the $(0^+, 1^+)$ doublet. We denote it
as $| 1^+, j_l={1\over 2}^+\rangle$. The other $1^+$ state $| 1^+,
j_l={3\over 2}^+\rangle$ belongs to $(1^+, 2^+)$ with $j_l={3\over
2}^+$. $j_l$ is a good quantum number when $m_Q\to \infty$.
Therefore $j_l=L+s_q={1\over 2}^+$ and $j_l=L+s_q={3\over 2}^+$
states don't mix with each other. They are exact mass eigenstates.
Moreover, it was shown that the mixing of these two states is
small even if one considers the $1/m_c$ correction in Ref.
\cite{zhu}. In other words, the physical state $D_{sj}(2460)$ is
quite close to $| 1^+, j_l={1\over 2}^+\rangle$.

In the quark model, the basis states are $^1P_1$ and $^3P_1$
states. Generally speaking, $D_{sj}(2460)$ should be the linear
combinations of $^1P_1$ and $^3P_1$ states \cite{1+mixing}
\begin{eqnarray}
| 1^+, j_l={1\over 2}^+\rangle &=&
\mathrm{cos\theta}\;|^1P_1\rangle+\mathrm{sin\theta}\;|^3P_1\rangle
\; ,\\
| 1^+, j_l={3\over 2}^+\rangle
&=&-\mathrm{sin\theta}\;|^1P_1\rangle+\mathrm{cos\theta}\;|^3P_1\rangle
\; .
\end{eqnarray}
The mixing angle $\theta_{HQ}=-\tan^{-1}\sqrt{2}$
 in heavy quark limit \cite{close}. With finite charm
quark mass, the mixing angle is not precisely known. However,
$\theta$ is not expected to differ dramatically from $-\tan^{-1}
\sqrt{2}$ according to Ref. \cite{zhu}.

\begin{table}[h]
\caption{The single-pion decay width (in keV) of $D_{sj}(2317,
2460)$ and $B_{sj}(5718, 5765)$.}\label{tab2}
\begin{center}
\begin{tabular}{  c  c l  }
\hline
Decay mode  &  & \\
 $A\rightarrow [B\pi^0]_L$& \raisebox{1.3ex}[0pt]{Amplitude} & \raisebox{1.3ex}[0pt]{$\Gamma$(keV)}\\
 \hline $D_{sj}(2317)\rightarrow [D_s \pi^0]_S$ &
 & 32  \\[0.5ex]
$B_s(0^+)\rightarrow [B_s(0^-) \pi^0]_S$& \raisebox{1.3ex}[0pt]{
$-{1\over48}W\Big{[}{-12\over
V}+{\mathbf{p^2}}\Big{(}1-{R^4_A\over V^2}\Big{)}\Big{]}$} & 35
\\ \hline $D_{sj}(2460)\to (D^*_s \pi^0)_S$ &
&  33($\theta_{HQ}$)\\
 $B_s(1^+)\rightarrow [B_s(1^-)
\pi^0]_S$&\raisebox{1.3ex}[0pt]{${1\over48\sqrt{3}}W
\Big{[}{-12\over V}+{\mathbf{p^2}} \Big{(}1-{R^4_A\over
V^2}\Big{)}\Big{]}(-\mathrm{cos\theta}+\sqrt{2}\mathrm{sin\theta})$}&38($\theta_{HQ}$)\\\hline
\end{tabular}
\end{center}
\end{table}

The the partial wave decay amplitude and width of $D_{sj}(2317,
2460)$ are listed in Table \ref{tab2}, where
\begin{eqnarray}
W&=&(2\pi)^{3\over2}\pi^{-{5\over4}}\;\gamma\delta_{\pi^0\eta}\;\Big{(}{R_A^5R^3_B
R^3_C}\Big{)}^{1\over2}\Big{(}{2\over V}\Big{)}^{3\over2} \sqrt{8
M_A E_B E_\pi}\exp\Big{(}-{1\over8}{\mathbf{p^2}}R_A^2{R_B^2+R_C^2\over V}\Big{)}\;,  \\
V &=&R_A^2+R_B^2+R_C^2\;.
\end{eqnarray}

\begin{figure}[hbt]
\begin{center}
\scalebox{0.8}{\includegraphics{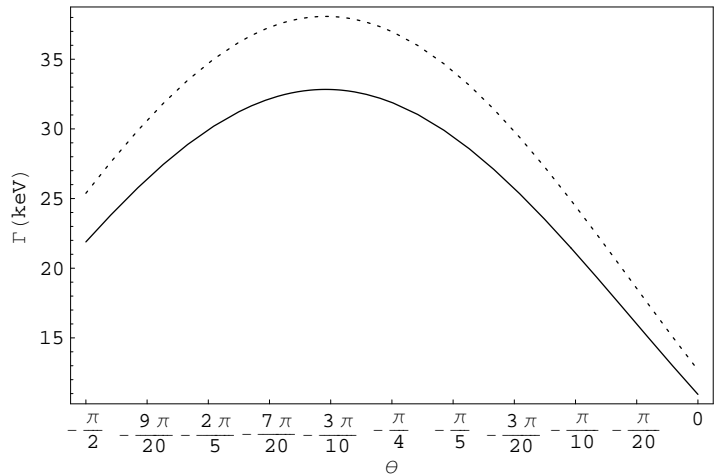}}
\end{center}
\caption{Variation of $B_s(5765)$'s (dashed curve) and
$D_{sj}(2460)$'s (solid curve) single pion S-wave decay width with
the mixing angle. }\label{fig3}
\end{figure}

The dependence of the single pion decay width of $D_{sj}(2317)$ on
the mixing angle is shown by Fig. \ref{fig3}. The mixing between
$^3P_1$ and $^1P_1$ states enhance $D_{sj}(2460)$'s single pion
S-wave decay width, and decreases the D-wave width. We ignore the
D-wave width because it is much smaller than the S-wave decay
width.

\section{Double pion decays of $D_{sj}(2460)$}\label{sec4}

Parity and angular momentum conservation forbid the double-pion
decay mode of $D_{sj}(2317)$ mesons. In contrast, $D_{sj}(2460)$
can decay into $D_s \pi\pi$ and $D_s^* \pi\pi$ final states. There
is no reliable way to calculate the multiple pion decay width. One
has to rely heavily on specific models to estimate the double-pion
decay width. For example, the double pion decay of $D_{sj}(2460)$
meson was assumed to occur via a virtual $\sigma$ meson in Ref.
\cite{Bardeen}.

\begin{table}[h]
\caption{The double-pion decay width (in keV) of $D_{sj}(2317,
2460)$ and $B_{sj}(5718, 5765)$. The subscript $A$ refers to
$D_{sj}(2460)$ and $B_s(1^+)$ respectively, $B$ refers to $D_s$,
$D^*_s$, $B_s(0^-)$ and $B_s(1^-)$ respectively, and $X$ refers to
the $f_0(980)$. } \label{tab3}
\begin{center}
\begin{tabular}{  c  c c c }
\hline
Decay mode  &  & \\
$A\to [Bf_0(980)]_L \to [B\pi\pi]_L$
&\raisebox{1.3ex}[0pt]{$\mathcal{V}\Big{(}A\to[Bf_0(980)]_L\Big{)}$}
& \raisebox{1.3ex}[0pt]{$\Gamma_1$}
&\raisebox{1.3ex}[0pt]{$\Gamma_2$}
 \\\hline
 $D_{sj}(2460)\rightarrow  [D_sf_0(980)]_P\rightarrow [D_s \pi\pi]_P$  &  &  2.6 & 6.9    \\
$B_s(0^+)\rightarrow[B_sf_0(980)]_P\rightarrow[B_s(0^-)\pi\pi]_P$
&\raisebox{1.3ex}[0pt]{${1\over4\sqrt{2}}W'\Big{[}\;4(2-5{R^2_A\over
V}) +(1-{R^4_A\over V^2})R^2_A\;{\bf{k_B}}^2
\;\Big{]}{\mathrm{cos}}\theta +W'{\mathrm{sin}}\theta$}
&0.45  & 1.0 \\
 \hline $D_{sj}(2460)\rightarrow  [D^*_sf_0(980)]_P\rightarrow [D^*_s \pi\pi]_P$ &
 &  0.055 & 0.13 \\
$B_s(1^+)\rightarrow[B_sf_0(980)]_P\rightarrow
[B_s(1^-)\pi\pi]_P$& \raisebox{1.3ex}[0pt]{
$W'{\mathrm{cos}}\theta
+{1\over4\sqrt{2}}W'\Big{[}\;4\Big{(}1-5{R^2_A\over V}\Big{)}
 +(1-{R^4_A\over V^2})R^2_A{\bf{k_B}}^2 \;\Big{]}{\mathrm{sin}}\theta$}
& 0.044   & 0.12 \\ \hline
\end{tabular}
\end{center}
\end{table}

\begin{figure}
\begin{center}
\scalebox{0.8}{\includegraphics{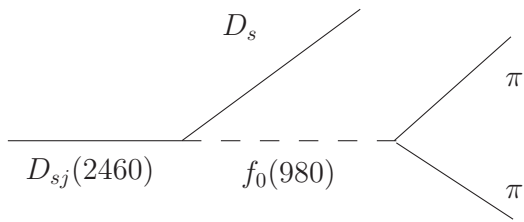}}
\end{center}
\caption{The double-pion decay modes via a virtual $f_0(980)$.
}\label{fig4}
\end{figure}

In our opinion, $f_0(980)$ is well established and less
controversial than the sigma meson. Moreover, its width is
measured to be 40-100 MeV \cite{pdg}. $f_0(980)$ mainly decays
into pions. This process can be described by the effective
Lagrangian
\begin{equation}
{\cal L}_{f_0(980)\pi\pi}=g_{f_0(980)\pi\pi} f_0 [2\pi^+\pi^-
+\pi^0\pi^0] \; .
\end{equation}
The coupling constant $g_{f_0\pi\pi}$ is $0.83\sim 1.3$ GeV. In
the following we assume the double pion decays occur through the
decay chain with the help of a virtual $f_0(980)$ meson
\footnote{We thank E. S. Swanson for helpful discussions on this
point.}: $D_{sj}(2460)\rightarrow D_s +f_0(980) \rightarrow D_s
+2\pi$ and $D_{sj}(2460) \rightarrow D^*_s +f_0(980) \rightarrow
D^*_s + 2\pi$. We further assume that the decay orbital angular
momentum \textbf{L} exists between $f_0(980)$ and $D_s$ (or
$D^\ast_s$) only. We want to remind the readers that the above
working assumption is only used to make a rough estimate of
$D_{sj}(2460)$'s double-pion width.

\begin{figure}
\begin{center}
\scalebox{0.6}{\includegraphics{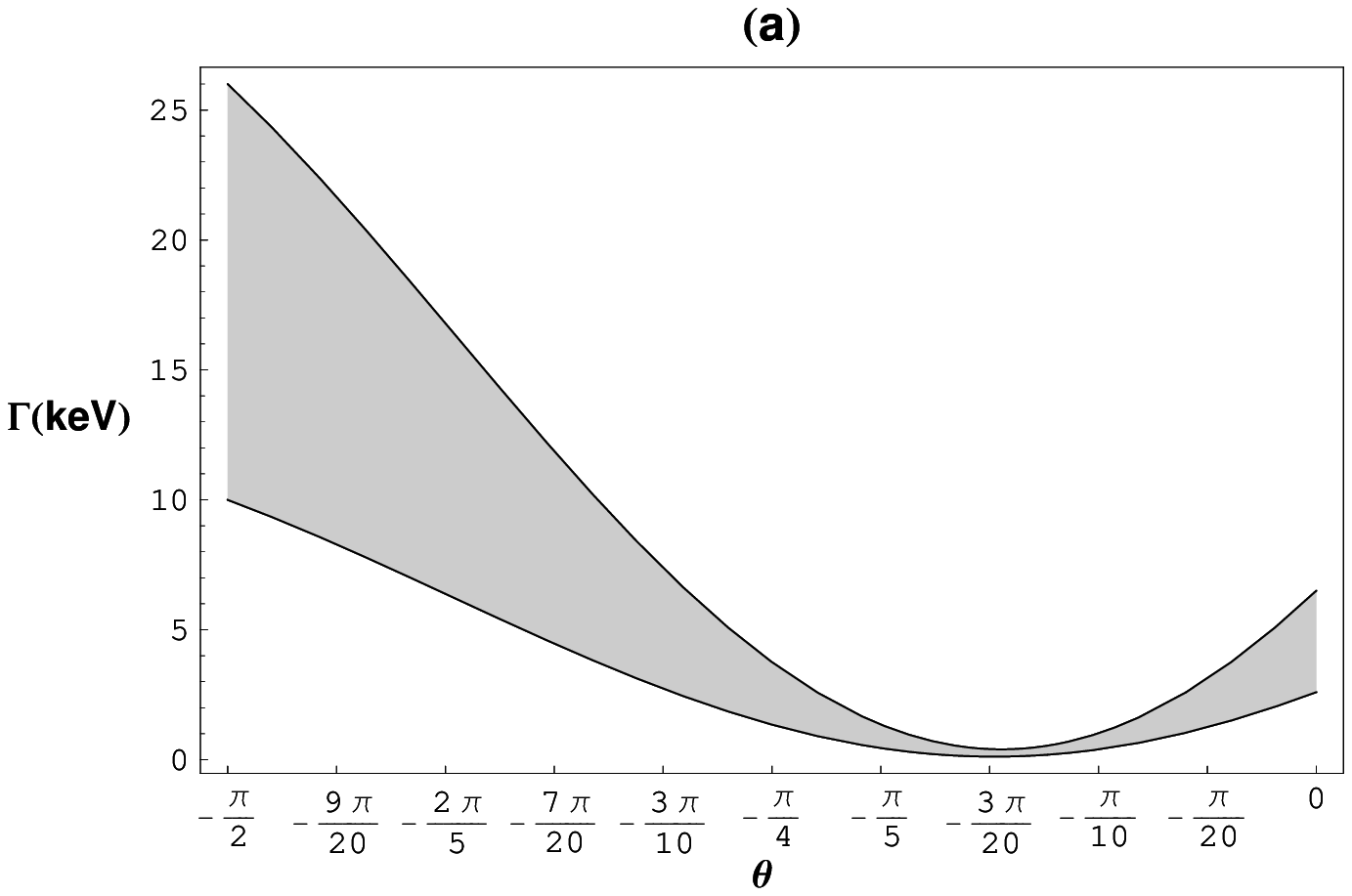}}\scalebox{0.59}{\includegraphics{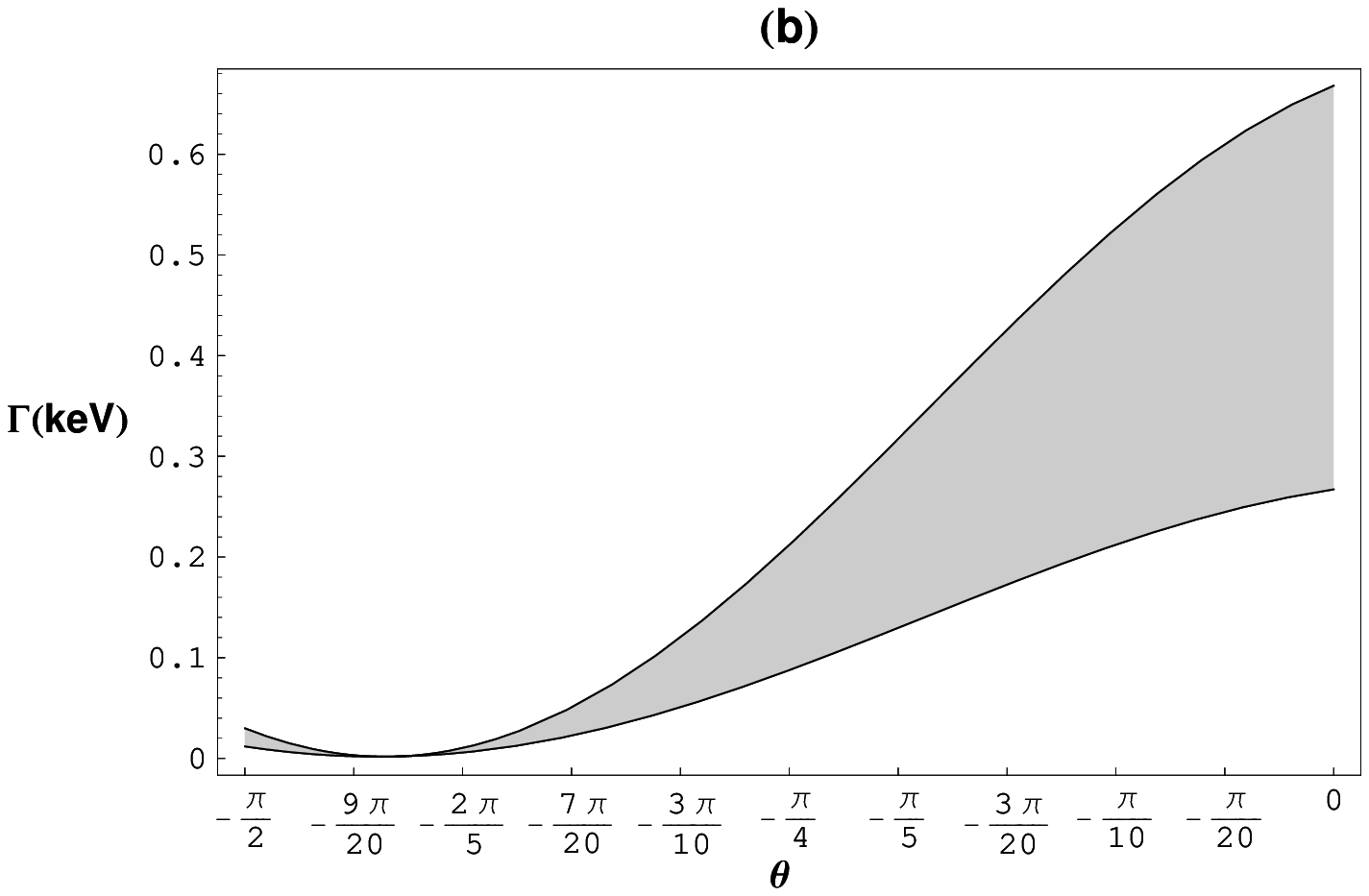}}
\end{center}
\caption{Variation of the double-pion decay width for the process
(a) $D_{sj}(2460)\to D_s \pi\pi $ and (b) $D_{sj}(2460)\to D^*_s
\pi\pi $ with $g_{f_0\pi\pi}=[0.59,0.93]$ GeV and mixing angle
$\theta$. The shadow is the possible value of
 the decay widths.}\label{1+mix2}
\end{figure}

We first consider the decay chain $D_{sj}(2460)\rightarrow D_s
+f_0(980) \rightarrow D_s +2\pi$. The decay amplitude is
\begin{eqnarray}\label{ee}
{\mathcal{M}}\Big{(}D_{sj}(2460)\rightarrow D_s + 2\pi\Big{)}=
{\mathcal{V}}\Big{(}D_{sj}(2460)\rightarrow D_s + f_0 \Big{)}
\cdot {i\over k^2_{f_0}-m^2_{f_0}} \cdot
\sqrt{\lambda_{\pi\pi}}\;g_{f_0 \pi \pi}
\end{eqnarray}
with ${\lambda_{\pi^+\pi^-}}=2$, ${\lambda_{\pi^0\pi^0}}=1$. The
vertex ${\mathcal{V}}\Big{(}D_{sj}(2460)\rightarrow D_s
+f_0\Big{)}$ can be obtained in $^3P_0$ model, which is the
function of momenta of $D_s's$ momentum $\mathbf{k_B}$. The
process $D_{sj}(2460) \rightarrow D^*_s +f_0(980)$ is Similar. We
list the amplitudes and widths of double pionic decays of
$D_{sj}(2460)$ and $B_{s}(5765)$ in Table \ref{tab3}, where
\begin{eqnarray*}
W'&=&(2\pi)^{3\over2}{2\sqrt{2}\over27}\pi^{1\over4}\;\gamma
\;\Big{(}R_A^5R_B^3R_X^5\Big{)}^{1\over2}\Big{(}{2\over
V}\Big{)}^{5\over2}\sqrt{8E_A E_B E_X}  \; |{\bf{k_B}}|\;
 \exp\Big{[}-{1\over 8}{\bf{k_B}}^2R_A^2{R_B^2+R_X^2\over
 V}\Big{]} \;.
\end{eqnarray*}
The dependence of their double pion decay widths with the mixing
angle is presented in Fig. \ref{1+mix2}.

For the pure academic purpose, we also estimate the doubly
isospin-violating P-wave decay: $D_{sj}(2460)\to
D_{sj}(2317)+\eta\to D_s(1869)+\eta+\eta\to
D_s(1869)+\pi^0+\pi^0$. Numerically the width of the above decay
mode is $10^{-6}$ eV, which is tiny and completely buried by the
isospin conserving double-pion decay mode.


\section{Discussion}

Assuming both $D_{sj}(2317)$  and $D_{sj}(2460)$ are ordinary
$c\bar s$ mesons, we have calculated their strong decay width
using the $^3P_0$ model. The single pion decays are
isospin-violating and occur through $\eta$-$\pi^0$ mixing. We have
collected the single pion decay widths available in literature in
Table \ref{tab4}. Our results agree very well with a recent
light-cone QCD sum rule calculation \cite{wei}.

\begin{table}[h]
\caption{\baselineskip 15pt Single-pion decay widths (in keV) of
$D_{sJ}(2317)$ and $D_{sJ}(2460)$ mesons from various theoretical
approaches.} \label{tab4}
\begin{center}
\begin{tabular}
{c c c c c c c c c c}
\hline  & this work& \cite{wei} &\cite{colangelo1} &
\cite{Bardeen} & \cite{godfrey} &\cite{fayy}&\cite{tetra1}&\cite{ishida}&\\
\hline
$D^*_{sJ}(2317)\rightarrow D_{s}\pi^0$& 32 &  34-44   & $7\pm 1  $& 21.5  &$\sim 10 $& 16 & 10-100&$150\pm 70$\\
$D_{sJ}(2460)\rightarrow D_{s}^{*}\pi^0$& 35 & 35-51  & $7\pm 1  $& 21.5  &$\sim 10 $& 32 &       &$150\pm 70$\\
\hline
\end{tabular}
\end{center}
\end{table}

The double pion decays of $D_{sj}(2460)$ are allowed by isospin
symmetry. But they are suppressed by three-body phase space. Under
a rather crude assumption that such decays occurs with the help of
a virtual $f_0(980)$ meson, we have estimated the double pion
decay width to be around $2.6\sim 6.9$ keV for $D_{sj}(2460) \to
D_s + 2\pi$ mode and $0.055\sim 0.13$ keV for $D_{sj}(2460) \to
D^*_s + 2\pi$ mode, depending on the total pionic width of
$f_0(980)$. The double pion decay width of $D_{sj}(2460)$ is
numerically much smaller than its single pion width because of the
cancellation from the mixing of $^3P_1$ and $^1P_1$ states.

Putting the single and double pion decay modes together, the
strong decay width of $D_{sj}(2317, 2460)$ is less than 50 keV.
Combining the radiative decay width, the total width of
$D_{sj}(2317, 2460)$ is less than 100 keV. Both resonances are
extremely narrow if they are $c\bar s$ states. A precise
measurement of their total widths may help distinguish theoretical
models of their quark content.

As an straightforward extension, we also calculate the pionic
widths of $B_{sj}$ mesons in the $(0^+, 1^+)$ doublet. With masses
estimated in Ref. \cite{Bardeen}, their single and double pion
decay widths are listed in Table \ref{tab3}.

\section*{Acknowledgments}

This project was supported by the National Natural Science
Foundation of China under Grants 10375003 and 10421503, Ministry
of Education of China, FANEDD, Key Grant Project of Chinese
Ministry of Education (NO 305001) and SRF for ROCS, SEM. J.L.
thanks Ying Yan for cross-checking the results and thanks E. S.
Swanson, Y. R. Liu, P. Z. Huang and F. L. Wang for useful
discussions.

\end{document}